\documentclass[11pt]{article}

\usepackage[a4paper,margin=3cm]{geometry}
\usepackage{amsmath,amsthm,amssymb,cite,graphicx}
\usepackage[tight]{subfigure}
\usepackage{algorithm}
\usepackage[noend]{algpseudocode}

\newcommand{\Input}{\item[\textbf{Input:}]}
\newcommand{\Output}{\item[\textbf{Output:}]}

\newcommand{\algo}[1]{\normalfont{\textsc{#1}}}
\newcommand{\gw}{\algo{GoeWill}}
\newcommand{\au}{\algo{ApxUndir}}
\newcommand{\ad}{\algo{ApxDir}}
\newcommand{\appalg}{\algo{Approx}}

\newcommand{\class}[1]{\ensuremath{\mathsf{#1}}}
\newcommand{\DP}{\class{P}}
\newcommand{\NP}{\class{NP}}
\newcommand{\APX}{\class{APX}}

\newcommand{\nat}{\ensuremath{\mathbb{N}}}
\newcommand{\integer}{\ensuremath{\mathbb{Z}}}

\newcommand{\unduniv}{{\mathcal{U}}}
\newcommand{\diruniv}{{\mathcal{D}}}
\newcommand{\oL}{\ensuremath{\overline{L}}}
\newcommand{\close}[1]{\langle #1 \rangle}

\newcommand{\prob}[1]{\textup{#1}}
\newcommand{\maxug}[1]{\prob{Max-$#1$-UCC}}
\newcommand{\maxdg}[1]{\prob{Max-$#1$-DCC}}
\newcommand{\minug}[1]{\prob{Min-$#1$-UCC}}
\newcommand{\mindg}[1]{\prob{Min-$#1$-DCC}}

\newtheorem{theorem}{Theorem}[section]
\newtheorem{lemma}[theorem]{Lemma}

\newenvironment{keywords}{\begin{quote} \small \textbf{Keywords:}}{\end{quote}}
\newenvironment{acmclass}{\begin{quote} \small \textbf{ACM Computing
   Classification:}}{\end{quote}}

\newcommand{\bemph}[1]{\textbf{\boldmath #1}}
\newcommand{\apx}{{\operatorname{apx}}}

\title{Minimum-weight Cycle~Covers and Their~Approximability%
\thanks{A preliminary version of this work will appear in the Proceedings of the 33rd
Workshop on Graph-Theoretic Concepts in Computer Science (WG 2006), Lecture
Notes in Computer Science.}}

\author{Bodo Manthey%
\thanks{Supported by the Postdoc-Program of the German Academic Exchange
Service (DAAD). On leave from Saarland University, Department of Computer
Science, P.~O.~Box 151150, 66041 Saarbr\"ucken, Germany.}}

\date{\small Yale University, Department of Computer Science \\
P.~O.~Box 208285, New Haven, CT 06520-8285, USA \\
\texttt{manthey@cs.yale.edu}}

\begin{document}
\maketitle

\begin{abstract}
  A cycle cover of a graph is a set of cycles such that every vertex is part of
  exactly one cycle. An $L$-cycle cover is a cycle cover in which the length of
  every cycle is in the set $L \subseteq \nat$.

  We investigate how well $L$-cycle covers of minimum weight can be
  approximated. For undirected graphs, we devise a polynomial-time approximation
  algorithm that achieves a constant approximation ratio for all sets $L$. On
  the other hand, we prove that the problem cannot be approximated within a
  factor of~$2-\varepsilon$ for certain sets $L$.

  For directed graphs, we present a polynomial-time approximation algorithm that
  achieves an approximation ratio of $O(n)$, where $n$ is the number of
  vertices. This is asymptotically optimal: We show that the problem cannot be
  approximated within a factor of $o(n)$.

  To contrast the results for cycle covers of minimum weight, we show that the
  problem of computing $L$-cycle covers of maximum weight can, at least in
  principle, be approximated arbitrarily well.
\end{abstract}


\begin{acmclass}
  F.2.2 [Analysis of Algorithms and Problem Complexity]:
     Nonnumerical Algorithms and Problems---computations on discrete structures;
  G.2.1 [Discrete Mathematics]: Combinatorics---combinatorial algorithms;
  G.2.2 [Discrete Mathematics]:
     Graph Theory---graph algorithms, path and circuit problems.
\end{acmclass}

\section{Introduction}
\label{sec:intro}

A cycle cover of a graph is a spanning subgraph that consists solely of cycles
such that every vertex is part of exactly one cycle. Cycle covers are an
important tool for the design of approximation algorithms for different variants
of the traveling salesman problem~\cite{Blaeser:ATSPZeroOne:2004,BlaeserEA:ATSP:2006,
   BlaeserEA:MetricMaxATSP:2005,BoeckenhauerEA:SharpenedIPL:2000,
   ChandranRam:Parameterized:2007,ChenNagoya:MetricMaxTSP:2007,
   ChenEA:ImprovedMaxTSP:2005,KaplanEA:TSP:2005},
for the shortest common superstring problem from computational
biology~\cite{BlumEA:Superstrings:1994,Sweedyk:ApproximationSuperstring:1999},
and for vehicle routing problems~\cite{HassinRubinstein:VehicleRouting:2005}.

In contrast to Hamiltonian cycles, which are special cases of cycle covers,
cycle covers of minimum weight can be computed efficiently. This is exploited in
the above mentioned algorithms, which in general start by computing a cycle
cover  and then join cycles to obtain a Hamiltonian cycle (this technique is
called \emph{subtour patching}~\cite{GilmoreEA:WellSolved:1985}).

Short cycles limit the approximation ratios achieved by such algorithms. Roughly
speaking, the longer the cycles in the initial cover, the better the
approximation ratio. Thus, we are interested in computing cycle covers without
short cycles. Moreover, there are algorithms that perform particularly well if
the cycle covers computed do not contain cycles of odd
length~\cite{BlaeserEA:ATSP:2006}. Finally, some vehicle routing
problems~\cite{HassinRubinstein:VehicleRouting:2005} require covering vertices
with cycles of bounded length.

Therefore, we consider \emph{restricted cycle covers}, where cycles of certain
lengths are ruled out a priori: For a set $L \subseteq \nat$, an
\emph{$L$-cycle cover} is a cycle cover in which the length of each cycle is
in~$L$.

Unfortunately, computing $L$-cycle covers is hard for almost all sets
$L$~\cite{HellEA:RestrictedTwoFactors:1988,Manthey:RestrictedCCWAOA:2006,
   Manthey:RestrictedCC:2007ECCC}.
Thus, in order to fathom the possibility of designing approximation algorithms
based on computing cycle covers, our aim is to find out how well $L$-cycle
covers can be approximated.

Beyond being a basic tool for approximation algorithms, cycle covers are
interesting in their own right. Matching theory and graph factorization are
important topics in graph theory. The classical matching problem is the
problem of finding one-factors, i.~e., spanning subgraphs in which every vertex
is incident to exactly one edge. Cycle covers of undirected graphs are also
called two-factors since every vertex is incident to exactly two edges in a
cycle cover. Both structural properties of graph factors and the complexity of
finding graph factors have been the topic of a considerable amount of research
(cf.\ Lov{\'a}sz and Plummer~\cite{LovaszPlummer:Matching:1986} and
Schrijver~\cite{Schrijver:CombOpt:2003}).

\subsection{Preliminaries}
\label{ssec:prelim}

Let $G=(V,E)$ be a graph with vertex set $V$ and edge set $E$. If $G$ is
undirected, then a \bemph{cycle cover} of $G$ is a subset $C \subseteq E$ of the
edges of $G$ such that all vertices in $V$ are incident to exactly two edges in
$C$. If $G$ is a directed graph, then a cycle cover of $G$ is a subset
$C \subseteq E$ such that all vertices are incident to exactly one incoming and
one outgoing edge in $C$. Thus, the graph $(V,C)$ consists solely of
vertex-disjoint cycles. The length of a cycle is the number of edges it consists
of. We are concerned with simple graphs, i.~e., the graphs do not contain
multiple edges or loops. Thus, the shortest cycles of undirected and directed
graphs are of length three and two, respectively. We call a cycle of length
$\lambda$ a \bemph{$\lambda$-cycle} for short.

An \bemph{$L$-cycle cover} of an undirected graph is a cycle cover in which the
length of every cycle is in the set $L \subseteq \unduniv = \{3,4,5,\ldots\}$.
An $L$-cycle cover of a directed graph is analogously defined except that
$L \subseteq \diruniv = \{2,3,4,\ldots\}$. A special case of $L$-cycle covers
are \bemph{$k$-cycle covers}, which are $\{k, k+1, \ldots\}$-cycle covers. Let
$\oL = \unduniv \setminus L$ in the case of undirected graphs, and let
$\oL = \diruniv \setminus L$ in the case of directed graphs (whether we consider
undirected or directed cycle covers will be clear from the context).

Given edge weights $w: E \rightarrow \nat$, the \bemph{weight $w(C)$} of a
subset $C \subseteq E$ of the edges of $G$ is $w(C) = \sum_{e \in C} w(e)$. In
particular, this defines the weight of a cycle cover since we view cycle covers
as sets of edges.

\bemph{\minug L} is the following optimization problem: Given an undirected
complete graph with non-negative edge weights that satisfy the triangle
inequality ($w(\{u,v\}) \leq w(\{u,x\})+w(\{x,v\})$ for all $u,x,v \in V$) find
an $L$-cycle cover of minimum weight. \bemph{\minug{k}} is defined for
$k \in \unduniv$ like \minug{L} except that $k$-cycle covers rather than
$L$-cycle covers are sought. The triangle inequality is not only a natural
restriction, it is also necessary: If finding $L$-cycle covers in graphs is
\NP-hard, then \minug L\ without the triangle inequality does not allow for any
approximation at all.

\bemph{\mindg L} and \bemph{\mindg k} are defined for directed graphs like
\minug{L} and \minug{k} for undirected graphs except that $L \subseteq \diruniv$
and $k \in \diruniv$ and the triangle inequality is of the form
$w(u,v) \leq w(u,x) + w(x,v)$.

Finally, \bemph{\maxug L}, \bemph{\maxug k}, \bemph{\maxdg L}, and
\bemph{\maxdg k} are analogously defined except that cycle covers of maximum
weight are sought and that the edge weights do not have to fulfill the triangle
inequality.

\subsection{Previous Results}
\label{ssec:previous}

\paragraph{Undirected Cycle Covers.}

\minug{\unduniv}, i.~e., the undirected cycle cover problem without any
restrictions, can be solved in polynomial time via Tutte's reduction to the
classical perfect matching problem~\cite{LovaszPlummer:Matching:1986}. By a
modification of an algorithm of Hartvigsen~\cite{Hartvigsen:PhD:1984}, also
4-cycle covers of minimum weight in graphs with edge weights one and two can be
computed efficiently. For \minug{k} restricted to graphs with edge weights one
and two, there exists a factor $7/6$ approximation algorithm for all
$k$~\cite{BlaeserSiebert:CycleCovers:2001}. Hassin and
Rubinstein~\cite{HassinRubinstein:TrianglePacking:2006Erratum}
presented a randomized approximation algorithm for \maxug{\{3\}} that achieves
an approximation ratio of $83/43 + \epsilon$. \maxug{L} admits a factor $2$
approximation algorithm for arbitrary sets
$L$~\cite{Manthey:ImprovedCC:2006,Manthey:RestrictedCC:2007ECCC}.
Goemans and Williamson~\cite{GoemansWilliamson:ConstrainedForest:1995} showed
that \minug k and \minug{\{k\}} can be approximated within a factor of $4$.
\minug L is \NP-hard and \APX-hard if $\oL \not\subseteq \{3\}$, i.~e., for all
but a finite number of sets $L$~\cite{HellEA:RestrictedTwoFactors:1988,
   Manthey:RestrictedCCWAOA:2006,Manthey:RestrictedCC:2007ECCC,
   Vornberger:EasyHard:1980}.
This means that for almost all $L$, these problems are unlikely to possess
polynomial-time approximation schemes (PTAS, see Ausiello et
al.~\cite{AusielloEA:ComplApprox:1999} for a definition).

If \minug L is \NP-hard, then the triangle inequality is necessary for efficient
approximations of this problem; without the triangle inequality, \minug L cannot
be approximated at all.

\paragraph{Directed Cycle Covers.}

\mindg{\diruniv}, which is also known as the \emph{assignment problem}, can be
solved in polynomial time by a reduction to the minimum weight perfect matching
problem in bipartite graphs~\cite{AhujaEA:NetworkFlows:1993}. The only other $L$
for which \mindg{L} can be solved in polynomial time is $L = \{2\}$. For all
$L \subseteq \diruniv$ with $L \neq \{2\}$ and $L \neq \diruniv$, \mindg L and
\maxdg L are \APX-hard and \NP-hard, even if only two different edge weights are
allowed~\cite{Manthey:RestrictedCCWAOA:2006,Manthey:RestrictedCC:2007ECCC}.

There is a $4/3$ approximation algorithm for
\maxdg 3~\cite{BlaeserEA:MetricMaxATSP:2005} as well as for \mindg{k} for
$k \geq 3$ with the restriction that the only edge weights allowed are one and
two~\cite{BlaeserManthey:MWCC:2005}. \maxdg{L} can be approximated within a
factor of $8/3$ for all $L$~\cite{Manthey:RestrictedCC:2007ECCC}.

Analogously to \minug L, \mindg L cannot be approximated at all without the
triangle inequality.

\subsection{New Results}
\label{ssec:new}

While $L$-cycle covers of \emph{maximum} weight allow for constant factor
approximations, only little is known so far about the approximability of
computing $L$-cycle covers of \emph{minimum} weight. Our aim is to close this gap.

We present an approximation algorithm for \minug L that works for all sets
$L \subseteq \unduniv$ and achieves a constant approximation ratio
(Section~\ref{ssec:goewill}). Its running-time is $O(n^2 \log n)$.
On the other hand, we show that the problem cannot be approximated within a
factor of $2-\varepsilon$ for general $L$ (Section~\ref{ssec:undinapp}).

Our approximation algorithm for \mindg L achieves a ratio of $O(n)$, where $n$
is the number of vertices (Section~\ref{ssec:directedalg}). This is
asymptotically optimal: There exists sets $L$ for which no algorithm can
approximate \mindg L within a factor of $o(n)$ (Section~\ref{ssec:inappdir}).
Furthermore, we argue that \mindg L is harder to approximate than the other
three variants even for more ``natural'' sets $L$ than the sets used to show the
inapproximability (Section~\ref{ssec:directedremarks}).

Finally, to contrast our results for \minug L and \mindg L, we show that
\maxug L and \maxdg L can be approximated arbitrarily well at least in principle
(Section~\ref{sec:maxgood}).

\section{\boldmath Approximability of \minug L}
\label{sec:appund}

\subsection{\boldmath An Approximation Algorithm for \minug L}
\label{ssec:goewill}

The aim of this section is to devise an approximation algorithm for \minug L
that works for all sets $L \subseteq \unduniv$. The catch is that for most $L$
it is impossible to decide whether some cycle length is in $L$ since there are
uncountably many sets $L$: If, for instance, $L$ is not a recursive set, then
deciding whether a cycle cover is an $L$-cycle cover is impossible. One option
would be to restrict ourselves to sets $L$ such that the unary language
$\{1^\lambda \mid \lambda \in L\}$ is in \DP. For such $L$, \minug L and
\mindg L are \NP\ optimization problems (see Ausiello et
al.~\cite{AusielloEA:ComplApprox:1999} for a definition). Another possibility
for circumventing the problem would be to include the permitted cycle lengths in
the input. While such restrictions are mandatory if we want to compute optimum
solutions, they are not needed for our approximation algorithms.

A complete $n$-vertex graph contains an $L$-cycle cover as a spanning subgraph
if and only if there exist (not necessarily distinct) lengths
$\lambda_1, \ldots, \lambda_k \in L$ for some $k \in \nat$ with
$\sum_{i=1}^k \lambda_i = n$. We call such an $n$ \bemph{$L$-admissible} and
define $\close L = \{n \mid \text{$n$ is $L$-admissible}\}$. Although $L$ can be
arbitrarily complicated, $\close L$ always allows efficient membership testing
according to the following lemma.

\begin{lemma}[\mbox{Manthey~\cite[Lem.~3.1]{Manthey:RestrictedCC:2007ECCC}}]
\label{lem:finite}
  For all $L \subseteq \nat$, there exists a finite set $L' \subseteq L$ with
  $\close{L'} = \close L$.
\end{lemma}

Let $g_L$ be the greatest common divisor of all numbers in $L$. Then $\close L$
is a subset of the set of natural numbers divisible by $g_L$. The proof of
Lemma~\ref{lem:finite} shows that there exists a minimum $p_L \in \nat$ such
that $\eta g_L \in \close L$ for all $\eta > p_L$. The number $p_L$ is the
Frobenius number~\cite{RamirezAlfonsin:Frobenius:2006} of the set
$\{\lambda \mid g_L \lambda \in L\}$, which is $L$ scaled down by $g_L$. For
instance, if $L = \{8,10\}$, then $g_L =2$ and $p_L = 11$ since the Frobenius
number of $\{4,5\}$ is $11$.

In the following, it suffices to know such a finite set $L' \subseteq L$. The
$L$-cycle covers computed by our algorithm will in fact be $L'$-cycle covers. In
order to estimate the approximation ratio, this cycle cover will be compared to
an optimal $\close{L'}$-cycle cover. Since
$L' \subseteq L \subseteq \close{L'}$, every $L'$- or $L$-cycle cover is also a
$\close{L'}$-cycle cover. Thus, the weight of an optimal $\close{L'}$-cycle
cover provides a lower bound for the weight of both an optimal $L'$- and an
optimal $L$-cycle cover. For simplicity, we do not mention $L'$ in the
following. Instead, we assume that already $L$ is a finite set, and we compare
the weight of the $L$-cycle cover computed to the weight of an optimal
$\close L$-cycle cover to bound the approximation ratio.

Goemans and Williamson have presented a technique for approximating constrained
forest problems~\cite{GoemansWilliamson:ConstrainedForest:1995}, which we will
exploit. Let $G=(V,E)$ be an undirected graph, and let $w: E \rightarrow \nat$
be non-negative edge weights. Let $2^V$ denote the power set of $V$. A
function $f: 2^V \rightarrow \{0,1\}$ is called a \bemph{proper function} if it
satisfies
\begin{itemize}
  \item $f(S) = f(V \setminus S)$ for all $S \subseteq V$ (symmetry),
  \item if $A$ and $B$ are disjoint, then $f(A) = f(B) = 0$ implies 
    $f(A \cup B) = 0$ (disjointness), and 
  \item $f(V) = 0$.
\end{itemize}
The aim is to find a set $F$ of edges such that there is an edge connecting $S$
to $V \setminus S$ for all $S \subseteq V$ with $f(S) = 1$. (The name
``constrained forest problems'' comes from the fact that it suffices to consider
forests as solutions; cycles only increase the weight of a solution.) For
instance, the minimum spanning tree problem corresponds to the proper function
$f$ with $f(S) = 1$ for all $S$ with $\emptyset \subsetneq S \subsetneq V$.

Goemans and Williamson have presented an approximation
algorithm~\cite[Fig.~1]{GoemansWilliamson:ConstrainedForest:1995} for
constrained forest problems that are characterized by proper functions. We will
refer to their algorithm as \gw.

\begin{theorem}[\mbox{Goemans and
   Williamson~\cite[Thm.~2.4]{GoemansWilliamson:ConstrainedForest:1995}}]
\label{thm:gw}
  Let $\ell$ be the number of vertices $v$ with $f(\{v\}) = 1$. Then \gw\ is a
  $(2-\frac 2\ell)$-approximation for the constrained forest problem defined by
  a proper function $f$.
\end{theorem}

In particular, the function $f_L$ given by
\[
f_L(S) =
\left\{
  \begin{array}{ll}
    1 & \text{if $|S| \not\equiv 0 \pmod{g_L}$ and} \\ 
    0 & \text{if $|S| \equiv 0 \pmod{g_L}$}
  \end{array}
\right.
\]
is proper if $|V| = n$ is divisible by $g_L$. (If $n$ is not divisible by $g_L$,
then $G$ does not contain an $L$-cycle cover at all.) Given this function, a
solution is a forest $H=(V,F)$ such that the size of every connected component
of $H$ is a multiple of~$g_L$. In particular, if $g_L = 1$, then $f_L(S) = 0$ 
for all $S$, and an optimum solution are $n$ isolated vertices.

If the size of all components of the solution obtained are in $\close L$, we are
done: By duplicating all edges, we obtain Eulerian components. Then we construct
an $\close L$-cycle cover by traversing the Eulerian components and taking
shortcuts whenever we come to a vertex that we have already visited. Finally, we
divide each $\lambda$-cycle into paths of lengths
$\lambda_1-1, \ldots, \lambda_k-1$ for some $k$ such that
$\lambda_1+ \ldots+ \lambda_k = \lambda$ and $\lambda_i \in L$ for all $i$. By
connecting the respective endpoints of each path, we obtain cycles of lengths
$\lambda_1, \ldots, \lambda_k$. We perform this for all components to get an
$L$-cycle cover. A straightforward analysis yields an approximation ratio of
$8$. A more careful analysis shows that the actual ratio achieved is $4$. The
details for the special case of $L = \{k\}$ are spelled out by Goemans and
Williamson~\cite{GoemansWilliamson:ConstrainedForest:1995}.

However, this procedure does not work for general sets $L$ since the sizes of
some components may not be in $\close L$. This can happen if $p_L > 0$ (for
$L = \{k\}$, for which the algorithm works, we have $p_L = 0$). At the end of
this section, we argue why it seems to be difficult to generalize the approach
of Goemans and Williamson in order to obtain an approximation algorithm for
\minug L whose approximation ratio is independent of $L$.

In the following, our aim is to add edges to the forest $H=(V,E)$ output by \gw\
such that the size of each component is in $\close L$. This will lead to an
approximation algorithm for \minug L with a ratio of $4 \cdot (p_L+4)$, which is
constant for each $L$. Let $F^\ast$ denote the set of edges of a minimum-weight
forest such that the size of each component is in $\close L$. The set $F^\ast$
is a solution to $G$, $w$, and~$f_L$, but not necessarily an optimum solution.

By Theorem~\ref{thm:gw}, we have $w(F) \leq 2 \cdot w(F^\ast)$ since $w(F^\ast)$
is at least the weight of an optimum solution to $G$, $w$, and $f_L$. Let
$C=(V', F')$ be any connected component of $F$ with $|V'| \notin \close L$. The
optimum solution $F^\ast$ must contain an edge that connects $V'$ to
$V \setminus V'$. The weight of this edge is at least the weight of the
minimum-weight edge connecting $V'$ to $V \setminus V'$.

We will add edges until the sizes of all components is in $\close L$. Our
algorithm acts in phases as follows: Let $H=(V, F)$ be the graph at the
beginning of the current phase, and let $C_1, \ldots, C_a$ be its connected
components, where $V_i$ is the vertex set of $C_i$. We will construct a new graph
$\tilde H=(V, \tilde F)$ with $\tilde F \supseteq F$. Let $C_1, \ldots, C_b$ be
the connected components with $|V_i| \notin \close L$. We call these components
\bemph{illegal}. For $i \in \{1, \ldots, b\}$, let $e_i$ be the cheapest edge
connecting $V_i$ to $V\setminus V_i$. (Note that $e_i = e_j$ for $i \neq j$ is
allowed.)

We add all these edges to $F$ to obtain
$\tilde F = F \cup \{e_1, \ldots, e_b\}$. Since $e_i$ is the cheapest edge
connecting $V_i$ to $V \setminus V_i$, the graph $\tilde H = (V,\tilde F)$ is
a forest. (If some $e_i$ are not uniquely determined, cycles may occur. We can
avoid these cycles by discarding some of the $e_i$ to break the cycles. For the
sake of simplicity, we ignore this case in the following analysis.) If
$\tilde H$ still contains illegal components, we set $H$ to be $\tilde H$ and
iterate the procedure.

\begin{lemma}
  Let $F$ and $\tilde F$ be as described above. Then
  $w(\tilde F) \leq w(F) + 2\cdot w(F^\ast)$.
\end{lemma}

\begin{proof}
  We observe that $F^\ast$ contains at least one edge $e^\ast_i$ connecting
  $V_i$ to $V \setminus V_i$ for every $i \in \{1,\ldots, b\}$. If
  $e^\ast_i = e^\ast_j$ for $i \neq j$, then $e_k^\ast \neq e_i^\ast$ for all
  $k \neq i,j$. This means that every edge occurs at most twice among
  $e_1^\ast, \ldots, e_b^\ast$, which implies
  \[
  \sum_{i=1}^b w(e_i^\ast) \leq 2 \cdot w(F^\ast).
  \]
  By the choice of $e_i$, we have $w(e_i) \leq w(e_i^\ast)$. Putting everything
  together yields
  \[
       w(\tilde F)
  \leq w(F) + \sum_{i=1}^b w(e_i)
  \leq w(F) + \sum_{i=1}^b w(e_i^\ast)
  \leq w(F) + 2w(F^\ast).
  \]
\end{proof}

Let us bound the number of phases that are needed in the worst case.

\begin{lemma}
  After at most $\lfloor p_L/2\rfloor +1$ phases, $\tilde H$ does not contain
  any illegal components.
\end{lemma}

\begin{proof}
  In the beginning, all components of $H=(V,F)$ contain at least $g_L$ vertices.
  If $g_L \in L$, no phases are needed at all. Thus, we can assume that
  $\min(L) \geq 2g_L$.

  To bound the number of phases needed, we will estimate the size of the
  smallest illegal component. Consider any of the smallest illegal components
  before some phase $t$, and let $s$ be the number of its vertices. In phase
  $t$, this component will be connected either to another illegal component,
  which results in a component with a size of at least $2s$, or to a legal
  component, which results in a component with a size of at least $s+2g_L$.
  (It can happen that more than two illegal components are connected to a single
   component in one phase.)

  In either case, except for the first phase, the size of the smallest illegal
  component increases by at least $2g_L$ in every step. Thus, after at most
  $\lfloor p_L/2\rfloor +1$ phases, the size of every illegal component is at
  least $(p_L+1)g_L$. Hence, there are no more illegal components since
  components that consist of at least $(p_L +1)g_L$ vertices are not illegal.
\end{proof}

Eventually, we obtain a forest that consists solely of components whose sizes
are in $\close L$. We call this forest $\tilde H=(V, \tilde F)$. Then we proceed
as already described above: We duplicate each edge, thus obtaining Eulerian
components. After that, we take shortcuts to obtain an $\close L$-cycle cover.
Finally, we break edges and connect the endpoints of each path to obtain an
$L$-cycle cover. The weight of this $L$-cycle cover is at most $4 \cdot w(\tilde F)$.

Overall, we obtain \au\ (Algorithm~\ref{algo:undirected}) and the following theorem.

\begin{algorithm}[t]
\begin{algorithmic}[1]
\Input undirected complete graph $G = (V, E)$, $|V| = n$;
       edge weights $w: E \rightarrow \nat$ satisfying the triangle inequality
\Output an $L$-cycle cover $C^\apx$ of $G$ if $n$ is $L$-admissible,
        $\bot$ otherwise
\If{$n \notin \close L$}
   \State return $\bot$
\EndIf
\State run \gw\ using the function $f_L$ described in the text to obtain
       $H=(V,F)$
\While{the size of some connected components of $H$ is not in $\close L$}
   \State let $C_1, \ldots, C_a$ be the connected components of $H$, where
          $V_i$ is the vertex set of $C_i$; let $C_1, \ldots, C_b$ be its
          illegal components
   \State let $e_i$ be the lightest edge connecting $V_i$ to $V \setminus V_i$
   \State add $e_1, \ldots, e_b$ to $F$
   \While{$H$ contains cycles}
      \State remove one $e_i$ to break a cycle
   \EndWhile
\EndWhile
\State duplicate each edge to obtain a multi-graph consisting of Eulerian
       components 
\ForAll{components of the multi-graph}
   \State walk along an Eulerian cycle
   \State take shortcuts to obtain a Hamiltonian cycle
   \State discard edges to obtain a collection of paths, the number of vertices
          of each of which is in $L$
   \State connect the two endpoints of every path in order to obtain cycles
\EndFor
\State the union of all cycles constructed forms $C^\apx$; return $C^\apx$
\end{algorithmic}
\caption{\au.}
\label{algo:undirected}
\end{algorithm}

\begin{theorem}
\label{thm:au}
  For every $L \subseteq \unduniv$, \au\ is a factor $(4\cdot (p_L +4))$
  approximation algorithm for \minug L. Its running-time is
  $O(n^2 \log n)$.
\end{theorem}

\begin{proof}
  Let $C^\ast$ be a minimum-weight $\close L$-cycle cover. The weight of
  $\tilde F$ is bounded from above by
  \[
       w(\tilde F)
  \leq   \left(\left\lfloor \frac{p_L}2 \right\rfloor +1 \right) \cdot 2 \cdot w(F^\ast)
       + 2 \cdot w(F^\ast)
  \leq  \bigl(p_L+4\bigr) \cdot w(C^\ast).
  \]
  Combining this with $w(C^\apx) \leq 4 \cdot w(\tilde F)$ yields the
  approximation ratio.

  Executing \gw\ takes time $O(n^2 \log n)$. All other operations can be
  implemented to run in time $O(n^2)$.
\end{proof}

\begin{figure}
\centering
\subfigure[The graph.]{%
   \label{fig:tightgraph}\includegraphics{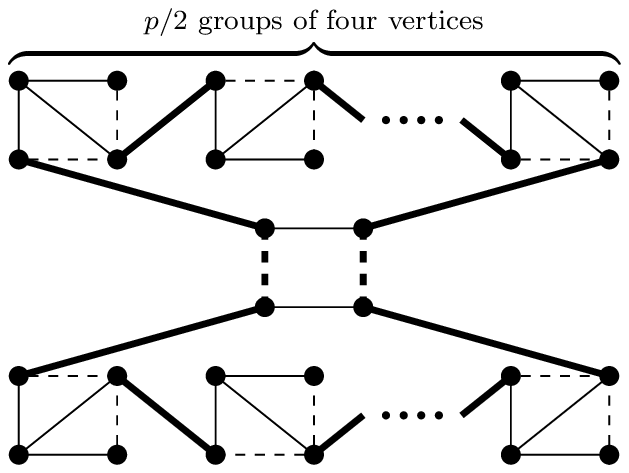}}
\qquad \quad
\subfigure[The optimal $L$-cycle cover.]{%
   \label{fig:tightoptimal}\includegraphics{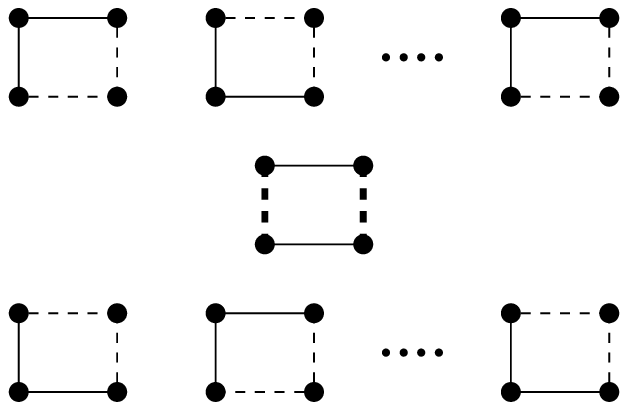}}
\caption{An example on which \au\ achieves only a ratio of roughly $p_L/2$.}
\label{fig:graphopt}
\end{figure}

We conclude the analysis of this algorithm by providing an example that shows
that the approximation ratio of the algorithm depends indeed linearly on $p_L$.
To do this, let $p \in \nat$ be even. We choose
$L = \{4, 2p+2, 2p+4, 2p+6, \ldots\}$. Thus, $g_L = 2$ and $p_L = p-1$.
Figure~\ref{fig:graphopt} shows the graph that we consider and its optimal
$L$-cycle cover. The graph consists of $4p + 4$
vertices. The weights of the edges, which satisfy the triangle inequality,
are as follows:
\begin{itemize}
   \item Solid, bold edges have a weight of $1$.
   \item Dashed, bold edges have a weight of $1+\varepsilon$, where
         $\varepsilon > 0$ can be made arbitrarily small.
   \item Solid, non-bold edges have a weight of $\varepsilon$.
   \item Dashed, non-bold edges have a weight of $2 \varepsilon$.
   \item The weight of the edges not drawn is given by the shortest path between
         the respective vertices.
\end{itemize}
The weight of the optimum $L$-cycle cover is
$2 + (6p+4)\varepsilon$: The four central
vertices contribute $2 + 4 \varepsilon$, and each of the $p$
remaining $4$-cycles contributes $6 \varepsilon$. By decreasing $\varepsilon$,
the weight of the optimum $L$-cycle cover can get arbitrarily close to $2$.

\begin{figure}
\centering
\subfigure[The output of \gw.]{%
   \label{fig:tightforest}\includegraphics{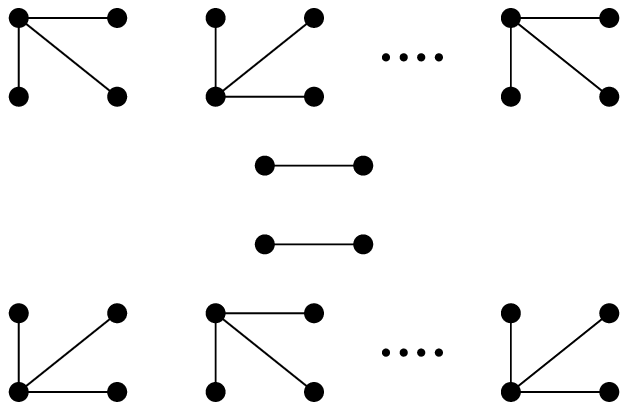}}
\qquad \quad
\subfigure[The final forest.]{%
   \label{fig:tightfinal}\includegraphics{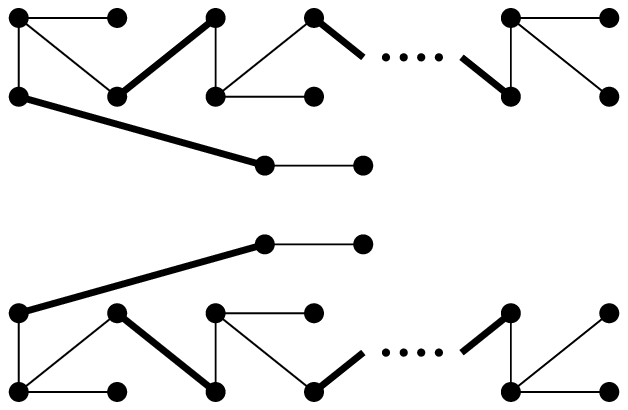}}
\\
\subfigure[The $L$-cycle cover $C^\apx$.]{%
   \label{fig:tightcycles}\includegraphics{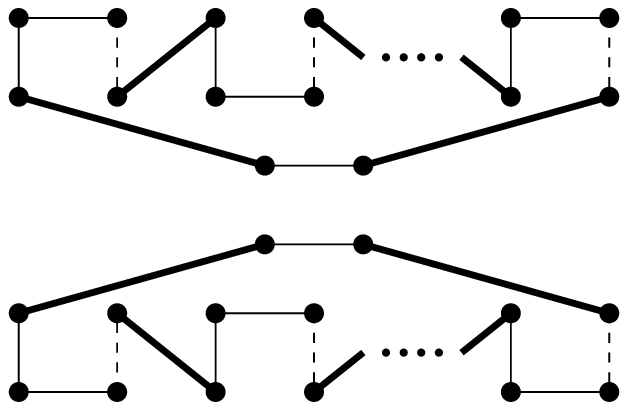}}
\caption{How \au\ computes an $L$-cycle cover of the graph of
   Figure~\ref{fig:tightgraph}.}
\label{fig:bad}
\end{figure}

Figure~\ref{fig:bad} shows what \au\ computes. Let us assume that \gw\ returns
the optimum $L$-forest shown in Figure~\ref{fig:tightforest}. \gw\ might also
return a different forest of the same weight: Instead of creating a component of
size four, it can take two vertical edges of weights $\varepsilon$ and
$2 \varepsilon$. However, the resulting $L$-cycle covers will be equal.

Starting with the output of \gw, \au\ chooses greedily the bold edges, which
have a weight of $1$, rather than the two edges of weight $1+\varepsilon$
(Figure~\ref{fig:tightfinal}). From the forest thus obtained, it constructs an
$L$-cycle cover (Figure~\ref{fig:tightcycles}). The weight of this $L$-cycle
cover is $2 (p/2 +1) + (4p +2)\varepsilon$. For sufficiently small
$\varepsilon$, this is approximately $p+2 = p_L+3$, which is roughly $p_L/2 + 3/2$
times as large as the weight of the optimum $L$-cycle cover.

Of course, it would be desirable to have an approximation algorithm with a ratio
that does not depend on $L$. Directly adapting the technique of Goemans and
Williamson~\cite{GoemansWilliamson:ConstrainedForest:1995} does not seem to
work: The function $f(S) = 1$ if and only if $|S| \notin \close L$ is not
proper because it violates symmetry. To force it to be symmetric, we can modify
it to $f'(S) = 1$ if and only if $|S| \notin \close L$ or
$|V \setminus S| \notin \close L$. But $f'$ does not satisfy disjointness. There
are generalizations of Goemans and Williamson's approximation technique to
larger classes of functions~\cite{GoemansWilliamson:PrimalDualNetwork:1997}.
However, it seems that $L$-cycle covers can hardly be modeled even by these more
general functions.

An alternative approach might be to grow a forest greedily without prior
execution of \gw. This works if $g_L = 1$. In this case, \gw\ outputs an empty
forest anyway since $f(S) = 0$ for all $S \subseteq V$, and \au\ boils down to a
greedy algorithm. However, if $g_L > 1$, then it is not guaranteed that we
obtain a feasible forest at all.

\subsection{\boldmath Unconditional Inapproximability of \minug L}
\label{ssec:undinapp}

In this section, we provide a lower bound for the approximability of \minug L as
a counterpart to the approximation algorithm of the previous section. We show that
the problem cannot be approximated within a factor of $2-\varepsilon$. This
inapproximability result is unconditional, i.~e., it does not rely on complexity
theoretic assumptions like $\DP \neq \NP$.

The key to the inapproximability of \minug L are
\bemph{immune sets}~\cite{Odifreddi:Recursion1:1989}: An infinite set
$L \subseteq \nat$ is called an immune set if $L$ does not contain an infinite
recursively enumerable subset. Such sets exist. One might want to argue that
inapproximability results based on immune sets are more of a theoretical
interest. But our result limits the possibility of designing general
approximation algorithms for $L$-cycle covers. To obtain algorithms with a ratio
better than 2, we have to design algorithms tailored to specific sets $L$.

Finite variations of immune sets are again immune sets. Thus for every
$k \in \nat$, there exist immune sets $L$ containing no number smaller than $k$.

\begin{theorem}
\label{thm:inappu}
  Let $\varepsilon > 0$ be arbitrarily small. Let $k > 2/\varepsilon$, and let
  $L \subseteq \{k, k+1, \ldots\}$ be an immune set. Then \minug{L} cannot be
  approximated within a factor of $2 - \varepsilon$.
\end{theorem}

\begin{proof}
  Let $G_n$ be an undirected complete graph with $n$ vertices
  $\{1,2, \ldots,n\}$. The weight of an edge $\{i,j\}$ for $i < j$ is
  $\min\{j-i, n+i-j\}$. This means that the vertices are ordered along an
  undirected cycle, and the distance from $i$ to $j$ is the number of edges that
  have to be traversed in order to get from $i$ to $j$. These edge weights
  fulfill the triangle inequality.

  For all $n \in L$, the optimal $L$-cycle cover of $G_n$ is a Hamiltonian cycle
  of weight $n$. Furthermore, the weight of every cycle $c$ that traverses
  $\ell \leq n/2$ vertices has a weight of at least $2\ell-2$: Let $i$ and $j$
  be two vertices of $c$ that are farthest apart according to the edge lengths
  of $G_n$. Assume that $i < j$. By the triangle inequality, the weight of $c$
  is at least $2 \cdot \min\{j-i, n+i-j\}$. Since $\ell \leq n/2$ and by the
  choice of $i$ and $j$, we have $\min\{j-i, n+i-j\} \geq \ell -1$, which proves
  $w(c) \geq 2\ell -2$.

  Consider any approximation algorithm \appalg\ for \minug{L}. We run \appalg\
  on $G_n$ for $n \in \nat$. By outputting the cycle lengths occurring in the
  $L$-cycle cover of $G_n$ for all $n$, we obtain an enumeration of a subset
  $S \subseteq L$. Since $L$ is immune, $S$ must be a finite set, and
  $s= \max(S)$ exists. Let
  $n \geq 2s$. The $L$-cycle cover output for $G_n$ consists of cycles whose
  lengths are at most $s \leq n/2$. Since $\min(L) \geq k$, we also have
  $\min(S) \geq k$ and the $L$-cycle cover output for $G_n$ consists of at most
  $n/k$ cycles. Hence, the weight of the cycle cover computed by \appalg\ is at
  least $\frac nk \cdot (2k-2)$. For $n \in L$, this is a factor of
  $2-\frac 2k > 2 - \varepsilon$ away from the optimum solution.
\end{proof}

Theorem~\ref{thm:inappu} is tight since $L$-cycle covers can be approximated
within a factor of $2$ by $L'$-cycle covers for every set $L' \subseteq L$ with
$\close{L'} = \close L$. For finite sets $L'$, all $L'$-cycle cover problems are
\NP\ optimization problems. This means that in principle optimum solutions can
be found, although this may take exponential time. The following
Theorem~\ref{thm:inappundtight} holds in particular for finite sets $L'$. In
order to actually get an approximation algorithm for \minug L out of it, we have
to solve \minug{L'} finite $L'$, which is \NP-hard and \APX-hard. But the proof
of Theorem~\ref{thm:inappundtight} shows also that any approximation algorithm
for \minug{L'} for finite sets $L'$ that achieves an approximation ratio of $r$
can be turned into an approximation algorithm for the general problem with a
ratio of $2r$.

Let $\min_L(G,w)$ denote the weight of a minimum-weight $L$-cycle cover of $G$
with edge weights $w$, which have to fulfill the triangle inequality.

\begin{theorem}
\label{thm:inappundtight}
  Let $L \subseteq \unduniv$ be a non-empty set, and let $L' \subseteq L$ with
  $\close{L'} = \close L$. Then we have
  $\min_{L'}(G,w) \leq 2 \cdot \min_{L}(G,w)$ for all undirected graphs $G$ with
  edge weights $w$ that satisfy the triangle inequality.
\end{theorem}

\begin{proof}
  Consider an arbitrary $L$-cycle cover $C$ and any of its cycles $c$ of length
  $\lambda \in L$. To prove the theorem, we show how to obtain an $L'$-cycle
  cover $C'$ from $C$ with $w(C') \leq 2 \cdot w(C)$. Consider any cycle $c$ of
  $C$ that has a length of $\lambda$. If $\lambda \in L'$, we simply put $c$
  into $C'$. Otherwise, since $\close{L'} = \close L \supseteq L$, there exist
  $\lambda_1, \ldots, \lambda_k \in L'$ for some $k \in \nat$ such that
  $\sum_{i=1}^k \lambda_i = \lambda$. We remove $k$ edges from $c$ to obtain $k$
  paths consisting of $\lambda_1, \ldots, \lambda_k$ vertices. No additional
  weight is incurred in this way. Then we connect the respective endpoints of
  each path to obtain $k$ cycles of lengths $\lambda_1, \ldots, \lambda_k$. By
  the triangle inequality, the weight of an edge added to close a cycle is at
  most the weight of the corresponding path. By performing this for every cycle
  of $C$, we obtain an $L'$-cycle cover $C'$ as claimed.
\end{proof}

\section{\boldmath Approximability of \mindg L}
\label{sec:inappdir}

\subsection{\boldmath An Approximation Algorithm for \mindg L}
\label{ssec:directedalg}

\begin{algorithm}[t]
\begin{algorithmic}[1]
\Input directed complete graph $G = (V, E)$, $|V| = n$;
       edge weights $w: E \rightarrow \nat$ satisfying the triangle inequality
\Output an $L$-cycle cover $C^\apx$ of $G$ if $n$ is $L$-admissible,
        $\bot$ otherwise
\If{$n \notin \close L$}
   \State return $\bot$
\EndIf
\State construct an undirected complete graph $G_U = (V,E_U)$ with edge weights
       $w_U(\{u,v\}) = w(u,v) + w(v,u)$
\State run \au\ on $G_U$ and $w_U$ to obtain $C_U^\apx$
\ForAll{cycles $c_U$ of $C_U^\apx$}
   \State $c_U$ corresponds to a cycle of $G$ that can be oriented in two ways;
          put the orientation $c$ that yields less weight into $C^\apx$
\EndFor
\State return $C^\apx$
\end{algorithmic}
\caption{\ad.}
\label{algo:directed}
\end{algorithm}

In this section, we present an approximation algorithm for \mindg L. The
algorithm exploits \au\ to achieve an approximation ratio of $O(n)$. The hidden
factor depends on $p_L$ again. This result matches asymptotically the lower
bound of Section~\ref{ssec:inappdir} and shows that \mindg L can be approximated
at least to some extent. (For instance, without the triangle inequality, no
polynomial-time algorithm achieves a ratio of $O(\exp(n))$ for an \NP-hard
$L$-cycle cover problem unless $\DP = \NP$.)

In order to approximate \mindg L, we reduce the problem to a variant of
\minug L, where also $2$-cycles are allowed: We obtain a $2$-cycle of an
undirected graph by taking an edge $\{u,v\}$ twice. Let $G=(V,E)$ be a directed
complete graph with $n$ vertices and edge weights $w:E \rightarrow \nat$ that
fulfill the triangle inequality. The corresponding undirected complete graph
$G_U = (V, E_U)$ has weights $w_U: E_U \rightarrow \nat$ with
$w_U(\{u,v\}) = w(u,v) + w(v,u)$.

Let $C$ be any cycle cover of $G$. The corresponding cycle cover $C_U$ of $G_U$
is given by $C_U = \{\{u,v\} \mid (u,v) \in C\}$. Note that we consider $C_U$ as
a multiset: If both $(u,v)$ and $(v,u)$ are in $C$, i.~e., $u$ and $v$ form a
$2$-cycle, then $\{u,v\}$ occurs twice in $C_U$. Let us bound the weight of $C_U$
in terms of the weight of $C$.

\begin{lemma}
\label{lem:nbound}
  For every cycle cover $C$ of $G$, we have $w_U(C_U) \leq n \cdot w(C)$.
\end{lemma}

\begin{proof}
  Consider any edge $e= (u,v) \in C$, and let $c$ be the cycle of length
  $\lambda$ that contains $e$. By the triangle inequality, we have
  $w_U(\{u,v\}) = w(u,v) + w(v,u) \leq w(c)$. Let $c_U$ be the cycle of $C_U$
  that corresponds to $c$. Since $c$ consists of $\lambda$ edges, we obtain
  $w_U(c_U) \leq \lambda \cdot w(c) \leq n \cdot w(c)$. Summing over all cycles
  of $C$ completes the proof.
\end{proof}

Our algorithm computes an $L'$-cycle cover for some finite $L' \subseteq L$
with $\close{L'} = \close L$. As in Section~\ref{ssec:goewill}, the weight of
the cycle cover computed is compared to an optimum $\close L$-cycle
cover rather than an optimum $L$-cycle cover. Thus, we can again assume that
already $L$ is a finite set.

The algorithm \au, which was designed for undirected graphs, remains to be an
$O(1)$ approximation if we allow $2 \in L$. The numbers $p_L$ and $g_L$ are
defined in the same way as in Section~\ref{ssec:goewill}.

Let $C_U^\apx$ be the $L$-cycle cover output by \au\ on $G_U$. We transfer
$C_U^\apx$ into an $L$-cycle cover $C^\apx$ of $G$. For every cycle $c_U$ of
$C_U^\apx$, we can orient the corresponding directed cycle $c$ in two
directions. We take the orientation that yields less weight, thus
$w(C^\apx) \leq w_U(C_U^\apx)/2$. Overall, we obtain \ad\
(Algorithm~\ref{algo:directed}), which achieves an approximation ratio of
$O(n)$ for every $L$.

\begin{theorem}
\label{thm:algodirected}
  For every $L \subseteq \diruniv$, \ad\ is a factor $(2n \cdot (p_L +4))$
  approximation algorithm for \mindg L. Its running-time is
  $O(n^2 \log n)$.
\end{theorem}

\begin{proof}
  We start by estimating the approximation ratio. Theorem~\ref{thm:au} yields
  $w_U(C_U^\apx) \leq 4\cdot (p_L +4) \cdot w_U(C_U^\ast)$, where $C_U^\ast$ is
  an optimal $\close L$-cycle cover of $G_U$. Now consider an optimum
  $\close L$-cycle cover $C^\ast$ of $G$. Lemma~\ref{lem:nbound} yields
  $w_U(C_U^\ast) \leq n \cdot w(C^\ast)$. Overall,
  \[
  w(C^\apx) \leq \frac 12 \cdot w_U(C_U^\apx)
            \leq 2\cdot (p_L +4) \cdot w_U(C_U^\ast) 
            \leq 2\cdot (p_L +4) \cdot n \cdot w(C^\ast).
  \]

  The running-time is dominated by the time needed to execute \gw\ in \au, which is
  $O(n^2 \log n)$.
\end{proof}

\subsection{\boldmath Unconditional Inapproximability of \mindg L}
\label{ssec:inappdir}

For undirected graphs, both $\maxug L$ and $\minug L$ can be approximated
efficiently to within constant factors. Surprisingly, in case of directed
graphs, this holds only for
the maximization variant of the directed $L$-cycle cover problem. \mindg L
cannot be approximated within a factor of $o(n)$ for certain sets $L$, where $n$
is the number of vertices of the input graph. In particular, \ad\ achieves
asymptotically optimal approximation ratios for \mindg L.

One might again want to argue that such an inapproximability result is more of
theoretical interest. But, similar to the case of \minug L, this result shows
that to find approximation algorithms, specific properties of the sets $L$ have
to be exploited. A general algorithm with a good approximation ratio for all
sets $L$ does not exist. Furthermore, as we will discuss in
Section~\ref{ssec:directedremarks}, \mindg L seems to be much harder a problem
than the other three variants, even for more practical sets~$L$.

\begin{theorem}
\label{thm:inapp}
  Let $L \subseteq \unduniv$ be an immune set. Then no approximation algorithm
  for \mindg L achieves an approximation ratio of $o(n)$, where $n$ is the
  number of vertices of the input graph.
\end{theorem}

\begin{proof}
  Let $G_n$ be a directed complete graph with $n$ vertices $\{1,2, \ldots,n\}$.
  The weight of an edge $(i,j)$ is $(j-i) \bmod n$. This means that the
  vertices are ordered along a directed cycle, and the distance from $i$ to $j$
  is the number of edges that have to be traversed in order to get from $i$ to
  $j$. These edge weights fulfill the triangle inequality.

  For all $n \in L$, the optimal $L$-cycle cover of $G_n$ is a Hamiltonian cycle
  of weight $n$. Furthermore, the weight of every cycle that traverses some of
  $G_n$'s vertices has a weight of at least $n$: Let $i$ and $j$ be two
  traversed vertices with $i<j$. By the triangle inequality, the path from $i$
  to $j$ has a weight of at least $j-i$ while the path from $j$ to $i$ has a
  weight of at least $i-j+n = (i-j) \bmod n$.

  Consider any approximation algorithm \appalg\ for \mindg L. We run \appalg\ on
  $G_n$ for $n \in \nat$. By outputting the cycle lengths occurring in the
  $L$-cycle cover of $G_n$ for all $n = 1, 2, \ldots$, we obtain an enumeration
  of a subset $S \subseteq L$.
  Since $L$ is immune, $S$ is a finite set, and $s = \max(S)$ exists. Thus, the
  $L$-cycle cover output for $G_n$ consists of at least $n/s$ cycles and has a
  weight of at least $n^2/s$. For $n \in L$, this is a factor of $n/s$ away from
  the optimum solution, where $s$ is a constant that depends only on \appalg.
  Thus, no recursive algorithm can achieve an approximation ratio of $o(n)$.
\end{proof}

\mindg{L'} for a finite set $L'$ is an \NP\ optimization problem. Thus, it can
be solved, although this may take exponential time. Therefore, the following
result shows that \mindg L can be approximated for all $L$ within
a ratio of~$n/s$ for arbitrarily large constants $s$, although this may also
take exponential time. In this sense, Theorem~\ref{thm:inapp} is tight.
We will first prove a lemma, which we will also use to prove Theorem~\ref{thm:maxptas}.

\begin{lemma}
\label{lem:numbertheory}
  For every $L \subseteq \nat$ and every $s > 1$, there exists a finite set
  $L' \subseteq L$ with $\close{L'} = \close L$ and the following property:
  For every $\lambda \in L \setminus L'$, there exist
  $\lambda_1, \ldots, \lambda_z \in L'$ with $z \leq \lambda/s$ such that
  $\sum_{i=1}^z \lambda_i = \lambda$.
\end{lemma}

\begin{proof}
  If $L$ is finite, we simply choose $L' = L$. So we assume that $L$ is
  infinite. Let again $g_L$ denote the greatest common divisor of all numbers of
  $L$. Let us first describe how to proceed if $g_L \in L$. After that we deal
  with the case that $g_L \notin L$.

  Let $L' = \{\lambda \in L \mid \lambda \leq m\}$, and let $\ell \in L'$. If
  $m$ is sufficiently large, then $\close{L'} = \close L$ (this follows from
  the proof of
  Lemma~\ref{lem:finite}~\cite[Lem.~3.1]{Manthey:RestrictedCC:2007ECCC} and also
  implicitly from this proof). We will specify $\ell$ and $m$, which depend on
  $s$, later on.

  Let $\lambda \in L \setminus L'$. Thus, $\lambda > m$. Let
  $r = \bmod(\lambda, \ell)$. Since $\lambda$ and $\ell$ are divisible by
  $g_L$, also $r$ is divisible by $g_L$. Since $\lambda \notin L'$, we have to
  find $\lambda_1, \lambda_2, \ldots \in L'$ that add up to $\lambda$. We have
  $\lambda = \lfloor \lambda/\ell \rfloor \cdot \ell + (r/g_L) \cdot g_L$. Now
  we choose
  $\lambda_1 = \ldots = \lambda_{\lfloor \lambda/\ell \rfloor } = \ell$ and
  $\lambda_{\lfloor \lambda/\ell \rfloor +1} = \ldots = 
   \lambda_{\lfloor \lambda/\ell \rfloor + r/g_L} = g_L$. What remains is to
  show that $\lfloor \lambda/\ell \rfloor + r/g_L \leq \lambda/s$.
  To do this, we choose $\ell > s$. Since $r/g_L$ is bounded from above by
  $\ell/g_L$, which does not depend on $\lambda$, we obtain
  $\lfloor \lambda/\ell \rfloor + r/g_L \leq \lambda/s$ for all $\lambda > m$
  for some sufficiently large $m$.

  The case that $g_L \notin L$ remains to be considered. There exist
  $\pi_1, \ldots, \pi_p \in L$ and $\xi_1, \ldots, \xi_p \in \integer$ for some
  $p \in \nat$ with $g_L = \sum_{i=1}^p \xi_i \pi_i$. Without loss of
  generality, we assume that $\xi_1 = \min_{1 \leq i \leq p} \xi_i$. We have
  $\xi_1 < 0$ since $g_L \notin L$.

  As above, let $L' = \{\lambda \in L \mid \lambda \leq m\}$, and let
  $\ell \in L'$. Let $\ell^\ast = -\xi_1 \ell \cdot \sum_{i=1}^p \pi_i > 0$. We
  choose $m$ to be larger than $\ell^\ast$. Let $\lambda > m$, and let
  $r = \bmod(\lambda - \ell^\ast,\ell)$. Then
  \begin{eqnarray*}
  \lambda  & = &  
     \left\lfloor \frac{\lambda - \ell^\ast}{\ell} \right\rfloor \cdot \ell
   + r + \ell^\ast \: = \:  \left\lfloor \frac{\lambda - \ell^\ast}{\ell} \right\rfloor \cdot \ell
   + \frac r{g_L} \cdot \sum_{i=1}^p \pi_i \xi_i - \xi_1 \ell 
                  \cdot \sum_{i=1}^p \pi_i \\ 
           & = &
     \left\lfloor \frac{\lambda - \ell^\ast}{\ell} \right\rfloor \cdot \ell
   + \sum_{i=1}^p \pi_i \cdot \left(\frac{r \xi_i}{g_L}  - \xi_1 \ell\right) .
  \end{eqnarray*}
  We have $\rho_i = \frac{r \xi_i}{g_L}  - \xi_1 \ell \geq 0$: Since $\xi_1 < 0$,
  we have $-\xi_1 \ell> 0$. If $\xi_i > 0$, then of course $\rho_i \geq 0$. If
  $\xi_i<0$, then $-\xi_i \leq -\xi_1$, and $\rho_i \geq 0$ follows from
  $r < \ell$.
  According to the deliberations above, we choose $\lambda_1 = \ldots =
  \lambda_{\lfloor (\lambda - \ell^\ast)/\ell \rfloor} = \ell$. In addition, we
  set $\rho_i$ of the $\lambda_j$s to $\pi_i$ for $1 \leq i \leq p$.
  It remains to be shown that $\lfloor (\lambda - \ell^\ast)/\ell \rfloor
  + \sum_{i=1}^p \rho_i \leq \lambda/s$. This follows from the fact that
  $\rho_i \leq \ell \cdot (\xi_i/g_L - \xi_1)$ for all $i$, which is independent
  of $\lambda$. Again, we choose $\ell > s$ and $m$ sufficiently large to
  complete the proof.
\end{proof}

\begin{theorem}
\label{thm:dirtight}
  For every $L$ and every $s > 1$, there exists a finite set $L' \subseteq L$ 
  with $\close{L'} = \close L$ such that
  $\min_{L'}(G,w) \leq \frac ns \cdot \min_{L}(G,w)$ for all directed graphs $G$
  with edge weights $w$.
\end{theorem}

\begin{proof}
  Let $s > 1$ and $L \subseteq \diruniv$ be given. We choose $L'\subseteq L$
  as described in the proof of Lemma~\ref{lem:numbertheory}. In order to prove
  the theorem, let $G$ be a directed complete graph, and let $C$ be an $L$-cycle
  cover of minimum weight of $G$. We show that we can find an $L'$-cycle cover
  $C'$ with $w(C') \leq \frac ns \cdot w(C)$.

  The $L'$-cycle cover $C'$ contains all cycles of $C$ whose lengths are in
  $L'$. Now consider any cycle $c$ of length $\lambda \in L \setminus L'$.
  According to Lemma~\ref{lem:numbertheory}, there exist
  $\lambda_1, \ldots, \lambda_z \in L'$ with $\sum_{i=1}^z \lambda_i = \lambda$
  and $z \leq \lambda/s$. We decompose $c$ into $z$ cycles of length
  $\lambda_1, \ldots, \lambda_z$. By the triangle inequality, the
  weight of each of these new cycles is at most $w(c)$. Thus, the total weight
  of all $z$ cycles is at most
  $z \cdot w(c) \leq (\lambda/s) \cdot w(c) \leq (n/s) \cdot w(c)$. By
  performing this for all cycles of $C$, we obtain an $L'$-cycle cover $C'$ with
  $\min_{L'}(G) \leq w(C') \leq (n/s) \cdot w(C) = (n/s) \cdot \min_L(G)$.
\end{proof}

\subsection{\boldmath Remarks on the Approximability of \mindg L}
\label{ssec:directedremarks}

It might seem surprising that \mindg L is much harder to
approximate than \minug L or the maximization problems \maxug L and \maxdg L. In
the following, we give some reasons why \mindg L is more difficult than the
other three $L$-cycle cover problems. In particular, even for ``easy'' sets $L$,
for which membership testing can be done in polynomial time, it seems that
\mindg L is much harder to approximate than the other three variants.

Why is minimization harder than maximization? To get a good approximation ratio
in the case of maximization problems, it suffices to detect a few ``good'',
i.~e., heavy edges. If we have a decent fraction of the heaviest edges, their
total weight is already within a constant factor of the weight of an optimal
$L$-cycle cover.
In order to form an $L$-cycle cover, we have to connect the heavy edges using
other edges. These other edges might be of little weight, but they do not
decrease the weight that we have already obtained from the heavy edges.

Now consider the problem of finding cycle covers of minimum weight. It does not
suffice to detect a couple of ``good'', i.~e., light edges: Once we have
selected a couple of good edges, we might have to connect them with heavy-weight
edges. These heavy-weight edges can worsen the approximation ratio dramatically.

Why is \mindg L harder than \minug L? If we have a cycle in an undirected graph
whose length is in $\close L$ but not in $L$ (or not in $L'$ but we do not know
whether it is in $L$), then we can decompose it into smaller cycles all lengths
of which are in $L$. This can be done such that the weight at most doubles (see
Section~\ref{sec:appund}). However, this does not work for directed cycles as we
have seen in the proof of Theorem~\ref{thm:inapp}: By decomposing a long cycle
into smaller ones, the weight can increase tremendously.

Finally, a question that arises naturally is whether we can do better if all
allowed cycle lengths are known a priori. This can be achieved by restricting
ourselves to sets $L$ that allow efficient membership testing. Another option is
to include the allowed cycle lengths in the input, i.~e., in addition to an
$n$-vertex graph and edge weights, we are given a subset of $\{2,3,\ldots, n\}$
of allowed cycle lengths.

The cycle cover problem with cycle lengths included in the input contains the
ATSP as a special case: for an $n$-vertex graph, we allow only cycles of length
$n$. Any constant factor approximation for this variant would thus immediately
lead to a constant factor approximation for the ATSP. Despite a considerable
amount of research devoted to the ATSP in the past decades, no such algorithm
has been found yet. This is an indication that finding a constant factor
approximation for the more general problem of computing directed cycle covers
might be difficult.

Now consider the restriction to sets $L$ for which
$\{1^\lambda \mid \lambda \in L\}$ is in $\DP$ (\mindg L is an \NP\ optimization
problem for all such $L$). If we had a factor $r$ approximation algorithm for
\mindg L for such $L$, where $r$ is independent of $L$, we would obtain a
$c \cdot \log n$ approximation algorithm for the ATSP, where $c> 0$ can be made
arbitrarily small: In particular, such an algorithm for \mindg L would allow for
an $r$ approximation of \mindg k for all $k \in \nat$. A close look at the
$(\log n)$ approximation algorithm for ATSP of Frieze et
al.~\cite{FriezeEA:TSP:1982} shows that an $r$-approximation for $k$-cycle
covers would yield an $(r \cdot \log_kn)$ approximation for the ATSP. We have
$r \cdot \log_kn = \frac{r}{\log k} \cdot \log n$. Thus, by increasing $k$, we
can make $c = \frac{r}{\log k}$ arbitrarily small. This would improve
dramatically over the currently best approximation ratio of
$0.842 \cdot \log_2 n$~\cite{KaplanEA:TSP:2005}.

\section{\boldmath Properties of Maximum-weight Cycle Covers}
\label{sec:maxgood}

To contrast our results for \minug L and \mindg L, we show that their
maximization counterparts \maxug L and \maxdg L can, at least in principle, be
approximated arbitrarily well; their inapproximability is solely due to their
\APX-hardness and not to the difficulties arising from undecidable sets $L$.
In other words, the lower bounds for \minug L and \mindg L presented in this
paper are based on the hardness of deciding whether certain lengths are in $L$.
The inapproximability of \maxug L and \maxdg L is based on the difficulty of
finding good $L$-cycle covers rather than testing whether they are $L$-cycle
covers.

Let $\max_L(G,w)$ be the weight of a maximum-weight $L$-cycle cover of $G$ with
edge weights $w$. The edge weights $w$ do not have to fulfill the triangle inequality.
We will show that $\max_L(G,w)$ can be approximated arbitrarily well by 
$\max_{L'}(G,w)$ for finite sets $L' \subseteq L$ with $\close{L'} = \close L$.
Thus, any approximation
algorithm for \maxug{L'} or \maxdg{L'} for finite sets $L'$ immediately yields
an approximation algorithm for general sets $L$ with an only negligibly worse
approximation ratio.
The following theorem for directed cycle covers contains the case of 
undirected graphs as a special case.

\begin{theorem}
\label{thm:maxptas}
  Let $L \subseteq \diruniv$ be any non-empty set, and let $\varepsilon > 0$.
  Then there exists a finite subset $L' \subseteq L$ with
  $\close{L'} = \close L$ such that
  $\max_{L'}(G,w) \geq (1-\varepsilon) \cdot \max_L(G,w)$ for all graphs $G$
  with edge weights $w$.
\end{theorem}

\begin{proof}
  Let $\varepsilon> 0$ be given. We choose $s > 1$ with $1/s \leq \varepsilon$.
  According to Lemma~\ref{lem:numbertheory}, there exists a finite set
  $L' \subseteq L$ with $\close{L'} = \close L$ with the following property: For
  all $\lambda \in L \setminus L'$, there exist
  $\lambda_1, \ldots, \lambda_z \in L'$ for
  $z \leq \lambda/s \leq \varepsilon \lambda$ that sum up to $\lambda$. Let us
  compare $\max_{L'}(G)$ and $\max_L(G)$. Therefore, let $C$ be an optimum
  $L$-cycle cover. We show how to obtain an $L'$-cycle cover $C'$ from $C$.
  The $L'$-cycle cover $C'$ contains all cycles of $C$ whose lengths are in
  $L'$. Let us consider any cycle $c$ of length $\lambda \in L \setminus L'$.
  There exist $\lambda_1, \ldots, \lambda_z \in L'$ for some
  $z \leq \varepsilon \lambda$ that sum up to $\lambda$. We break $z$ edges of
  $c$ to obtain a collection of paths of lengths
  $\lambda_1-1, \ldots, \lambda_z-1$. Since we break at most an $\varepsilon$
  fraction of $c$'s edges, we can remove these $z$ edges such that at most an
  $\varepsilon$ fraction of $w(c)$ is lost. Then we connect the respective
  endpoints of each path to obtain $z$ cycles of lengths
  $\lambda_1, \ldots, \lambda_z$. No weight is lost in this way.

  We have lost at most $\varepsilon \cdot w(c)$ of the weight of every cycle $c$
  of $C$, thus $\max_{L'}(G) \geq w(C') \geq (1-\varepsilon) \cdot w(C) =
  (1-\varepsilon) \cdot \max_L(G)$.
\end{proof}

\section{Concluding Remarks}
\label{sec:concl}

First of all, we would like to know whether there is a general upper bound for
the approximability of \minug L: Does there exist an $r$ (independent of $L$)
such that \minug L can be approximated within a factor of $r$? We conjecture
that such an algorithm exists. If such an algorithm works also for the slightly
more general problem \minug L with $2 \in L$ (see
Section~\ref{ssec:directedalg}), then we would obtain a factor $rn/2$
approximation for $\mindg L$ as well.

While the problem of computing $L$-cycle cover of minimum weight can be
approximated efficiently in the case of undirected graphs, the directed variant
seems to be much harder. We are interested in developing approximation
algorithms for  \mindg L for particular sets $L$ or for certain classes of sets
$L$. For instance, how well can \mindg L be approximated if $L$ is a finite set?
Are there non-constant lower bounds for the approximability of \mindg L, for
instance bounds depending on $\max(L)$? Because of the similarities between
\mindg L and ATSP, an answer to either question would hopefully also shed some
light on the approximability of the ATSP.


\bibliographystyle{plain}
\bibliography{abbrev,bodo,graphs,tcs,approx,compbio,chapters,books,theses}

\end{document}